\documentclass[11pt,a4paper]{article}
\pdfoutput=1
\usepackage{jheppub}
\usepackage{amsmath}
\usepackage{verbatim}
\usepackage{graphicx}
\usepackage{mathrsfs}
\usepackage{appendix}
\usepackage{caption}
\usepackage{float}
\usepackage{subfig}
\usepackage{mathtools}
\usepackage{mathtools}

\usepackage{amssymb}
\usepackage{inputenc, array}
\usepackage{textcomp}
\usepackage[dvipsnames]{xcolor}

\newcommand{\z}{{\bar z}}
\newcommand{\h}{{\bar h}}

\renewcommand\O{{\mathcal{O}}}
\newcommand{\be}[1]{ \begin{equation}\label{#1} }
\newcommand{\ee}{\end{equation}}
\newcommand{\bea}[1]{\begin{eqnarray}\label{#1} }
\newcommand{\eea}{\end{eqnarray}}
\newcommand{\bes}{\begin{subequations}}
\newcommand{\ees}{\end{subequations}}

\newcommand{\p}{\partial}

\renewcommand{\(}{\left(}
\renewcommand{\)}{\right)}

\DeclarePairedDelimiterX\braket[2]{\langle}{\rangle}{#1 \delimsize\vert #2}

\renewcommand{\(}{\left(}
\renewcommand{\)}{\right)}

\title{Massive fields in 3D Minkowski space and boundary correlators}
\author{Sudipta Dutta} 

\affiliation{Université Libre de Bruxelles and International Solvay Institutes, Brussels, Belgium.}

\emailAdd{sudipta.dutta@ulb.be}

\abstract{A codimension-one Carrollian CFT on null infinity has been proposed as the putative dual description of asymptotically flat spacetimes and has so far been successful in describing the massless S-matrices in one higher dimension. In this work, we investigate the current proposal of Carrollian holography to include the massive  fields in the bulk. We discover a broader class of correlation functions in two-dimensional Carrollian CFTs and show that they encode massive scattering amplitudes in three-dimensional Minkowski spacetime. We also construct a bulk-to-boundary propagator that generalizes the one already existing for massless fields.}

\preprint{}

\begin{document}
\maketitle

\vfill

\newpage

\section{Introduction}

Generalizing the holographic principle to the physically relevant setting of asymptotically flat spacetimes is one of the outstanding open problems in theoretical physics. Unlike the AdS/CFT correspondence, there is currently no top-down construction of holography in Minkowski spacetime; consequently, most approaches are inherently bottom-up. In a bottom-up approach, one attempts to construct the holographic dual guided by robust features of the gravitational theory, such as its observables, symmetries, and representations. For asymptotically flat spacetimes, one such well-defined observable is the S-matrix. From the holographic viewpoint, the S-matrix plays a role analogous to boundary correlation functions in AdS/CFT. A successful holographic description of asymptotically flat spacetimes should therefore provide a boundary theory whose correlation functions encode the bulk S-matrix.
In addition, symmetry principles provide crucial constraints on the structure of the putative dual theory. The relevant symmetry group in this context is the isometry group of Minkowski spacetime, namely the Poincaré group, which in the presence of gravity is enhanced to an infinite-dimensional asymptotic symmetry group: the Bondi–van der Burg–Metzner–Sachs (BMS) group \cite{Bondi:1962px,Sachs:1962wk,Barnich:2009se,Campiglia:2014yka}. An important insight due to Strominger was that the infinite-dimensional BMS symmetries is also respected by the quantum gravitational S-matrix in asymptotically flat spacetimes. This symmetry completely controls the universal infrared behaviour of scattering amplitudes through the associated Ward identities and soft theorems \cite{Strominger:2013jfa,Strominger:2017zoo}. In particular, the leading and subleading soft graviton theorems are direct consequences of BMS invariance and can be understood as Ward identities of the associated charges \cite{He:2014laa,Kapec:2014opa}. This universal infrared behaviour of asymptotically flat gravity provided an organizing principle for constructing a holographic description of Minkowski spacetime. Depending on how the infinite-dimensional symmetry group is encoded in a lower-dimensional system, two major proposals have emerged: celestial holography \cite{Pasterski:2021rjz,Raclariu:2021zjz,McLoughlin:2022ljp,Donnay:2023mrd} and Carrollian holography \cite{Bagchi:2025vri,Nguyen:2025zhg,Ruzziconi:2026bix}. In this paper, we shall mainly focus on the Carrollian approach. \\

The Carrollian proposal originated from considering the large-radius limit of AdS/CFT. Group-theoretic arguments indicate that taking the flat-space limit in the bulk, corresponding to an infinite AdS radius or vanishing cosmological constant, induces a Carrollian ($c\to0$) contraction of the boundary CFT \cite{Bagchi:2010eg,Duval_2014}. From a geometric perspective, null infinity, intrinsically is a Carrollian manifold, and the action of the BMS group endows this intrinsic geometry with a conformal structure. Consequently, null infinity provides a natural arena for representations of the conformal Carroll group, leading to a codimension-one description of the putative holographic dual \footnote{The reader is pointed to \cite{Bagchi:2012xr,Barnich:2012xq,Barnich:2013yka,Barnich:2015mui,Bagchi:2014iea,Jiang:2017ecm,Hartong:2015usd,Detournay:2014fva,Bagchi:2015wna,Poulias:2025eck,Hao:2025btl,Cotler:2024cia,Aggarwal:2025hji} for discussions on holographic dual for three dimensional flat space in Carrollian approach.}. In this framework, the bulk scattering amplitudes are expected to be encoded in correlators of a Carrollian conformal field theory living on the entirety of null infinity \footnote{The reader is pointed to \cite{Donnay:2022aba,Bagchi:2022emh,Donnay:2022wvx,Bagchi:2023fbj,Nguyen:2023vfz,Nguyen:2023miw,Bagchi:2023cen,Salzer:2023jqv,Mason:2023mti,deBoer:2021jej,Saha:2023hsl,Bagchi:2024gnn,Ruzziconi:2024zkr,Kim:2023qbl,Kraus:2024gso,Kraus:2025wgi,Isen:2026xoc,Stieberger:2024shv,Liu:2024nfc,Kulp:2024scx,Chakrabortty:2024bvm,Navarro:2025xln,Berenstein:2025tts,Bagchi:2024efs,Lipstein:2025jfj,Adamo:2025bfr,Ammon:2025avo,Opreij:2026bdx,Nenmeli:2026ket,Long:2026xsf,Bagchi:2026emg} for discussions on Carrollian approach towards S-matrix}. Such a relation between massless scattering amplitudes and the primary Carroll correlators was found in \cite{Banerjee:2018gce,Bagchi:2022emh,Donnay:2022wvx}. This is operationally obtained by going into a basis that diagonalises the Lorentz boost generator instead of the translations. The integral transformation required to make this change of basis, is in essence is equivalent to the HKLL formula of AdS/CFT, and can alternatively be derived using the expression of the bulk to boundary propagator for Minkowski space. This propagator for massless fields was obtained independently using several methods, which includes, limit from AdS \footnote{\cite{PipolodeGioia:2022exe} obatains analogous result for celestial CFT implementing a large radius limit.} \cite{Bagchi:2023fbj}, an extrapolation to null infinity \cite{Alday:2024yyj}, and a path integral formulation of S-matrices \cite{Kraus:2024gso}.

\smallskip

\bigskip

\medskip
The goal of this paper is to extend this framework  to incorporate massive fields. In particular, we consider a simple set-up of a massive scalar field in $(2+1)$ dimensional Minkowski space. We find the bulk to boundary propagator using the massive Feynmann propagator and using a suitable extrapolation to null infinity \footnote{Extrapolation of massive fields to timelike infinity was discussed in \cite{Have:2024dff}.}. We also show the result obtained, is can infact be uniquely determined by symmetry arguments. We compute correlation functions using these propagators, which results to primary correlators of a codimension one Carrollian CFTs. This formulation yields a  relation between the massive scattering amplitudes and the boundary Carroll correlators. \footnote{An alternate proposal for describing S-matrix as boundary correlator was put forward in \cite{Jain:2023fxc}}.

\section{Elements of 2d Carrollian CFT} \label{2}

The asymptotic symmetries of three dimensional Einstein gravity, the BMS$_3$ group is generated by a infinite tower of superrotations ($L_n$) and supertranslations ($M_n$). They satisfy
\begin{align}
[L_m,L_n]&=(m-n)L_{m+n}++\frac{c_L}{12}(n^3-n)\delta_{n+m,0},\\ \nonumber
[L_m,M_n]&=(m-n)M_{m+n}+\frac{c_M}{12}(n^3-n)\delta_{n+m,0},\\ \nonumber
[M_m,M_n]&=0.
\end{align}
For Einstein gravity these central charges assume the values $c_L=0$, and $c_M=\frac{3}{G}$ \cite{Barnich:2006av}.
From the intrinsic perspective of null infinity $(\mathcal{I})$, these symmetry generators arise as the infinite-dimensional local conformal isometries. The superrotations $(L_n)$ generate the diffeomorphism of the celestial circle, while the supertranslations $(M_n)$ generate an angle-dependent translation along the null direction. The asymptotic generators, when projected on null infinity, take the following form
\begin{align}
    L_{n}=ie^{in\phi}(\p_{\phi}+inu\p_u), \quad M_{n}=ie^{in\phi}\p_u
\end{align}
They can alternatively be obtained by solving a conformal killing equation with respect to the intrinsic Carrollian `metric´ on null infinity.
Of particular importance is the finite-dimensional global subalgebra generated by $\{L_0,L_{\pm1},M_0,M_{\pm1}\}$
,which is isomorphic to the three-dimensional Poincaré algebra $\mathfrak{iso}(2,1)$. The explicit identifications are given by
\begin{equation} \label{isomorphism}
L_0=J_{12},\qquad
L_{\pm1}=J_{01}\pm iJ_{02},
\qquad
M_0=P_0,\qquad
M_{\pm1}=P_1\pm iP_2,
\end{equation}
where $J_{\mu\nu}$ and $P_\mu$ denote the Lorentz and translation generators respectively. Consequently, a conformal field theory that lives on $\mathcal{I}$ would inherit the conformal isometries of the background, thus would be BMS$_3$ or conformal Carroll invariant. The representations of 2d Carroll CFTs have extensively been studied in the literature. In what follows, we shall recollect some basic ingredients of Carrollian representations and compute the lower point correlation functions of the Primary operators.  \\

\subsection{Primary operators}

A primary operator in 2d Carrollian CFTs, $\mathcal O_{\Delta,\xi}(t,x)$ is characterized by \cite{Bagchi:2009pe} 
\begin{equation}
[L_0,\mathcal O_{\Delta,\xi}(0,0)]
=
\Delta\,\mathcal O_{\Delta,\xi}(0,0),
\qquad
[M_0,\mathcal O_{\Delta,\xi}(0,0)]
=
\xi\,\mathcal O_{\Delta,\xi}(0,0).
\end{equation}
Here $(t,x)$ are coordinates on a Carrollian plane ($\mathbb{R\times R}$), and the eigenvalues $\Delta$ and $\xi$ are, respectively, the scaling dimension and the boost charge. On an arbitrary point, the transformation rules are given by

\begin{subequations}
    \begin{align}
    [L_n,\mathcal O_{\Delta,\xi}(t,x)]=
\left[
x^{n+1}\partial_x
+
(n+1)x^n t\partial_t
+
(n+1)\Delta x^n
+
n(n+1)\xi x^{n-1}t
\right]
\mathcal O_{\Delta,\xi}(t,x)
\end{align}
\begin{align}
    [M_n,\mathcal O_{\Delta,\xi}(t,x)]=
\left[
x^{n+1}\partial_t
+
(n+1)\xi x^n
\right]
\mathcal O_{\Delta,\xi}(t,x)
\end{align}

\end{subequations}

However, as $\mathcal{I}$ has the topology of $\mathbb{R}\times \mathbb{S}^1$, we would be interested in Carrollian CFT on a ´cylinder`. A plane-to-cylinder map of the form $x=e^{i\phi}$, $t=i\tau e^{i\phi}$ could be used to obtain the associated transformation rules on the cylinder. They are given by
\begin{subequations} \label{primary}
    \begin{align}
        &[L_n,\mathcal O_{\Delta,\xi}(u,\phi)]=i e^{in\phi}
\left(
\partial_\phi
+in u\,\partial_u
+in\Delta
-n^2\xi\,u
\right)
\mathcal O_{\Delta,\xi}(u,\phi),
\end{align}
\begin{align}
&[M_n,\mathcal O_{\Delta,\xi}(u,\phi)]=
i e^{in\phi}
\left(
\partial_u
+in\xi
\right)
\mathcal O_{\Delta,\xi}(u,\phi),
\end{align}
\end{subequations}
 Here $(\tau,\phi)$ denotes the coordinates on the cylinder.  A quasi-primary operator, analogous to CFT$_2$, is defined by restricting the transformation rules for $n=0,\pm1$ only. \\
Before moving to the correlation functions, let us emphasize that the \textit{mass-squared} Casimir of the global Poincare subgroup, yields the following value on a primary operator
\begin{equation}
     C=P^2=-(M_0^2-M_1M_{-1}), \quad    [C,\mathcal O_{\Delta,\xi}(u,\phi)]=-\xi^2 \mathcal O_{\Delta,\xi}(u,\phi)
\end{equation}
One might expect from the above Casimir argument that a boundary primary operator at null infinity  would require to have a non-zero boost charge $\xi$ to holographically describe the dynamics of a massive field in the bulk. One example of such an operator is the analog of vertex operators in free Carrollian scalar model \cite{Hao:2021urq}. We shall verify this matching of quantum numbers in explicit calculation of correlation functions in section \ref{4}.

\subsection{Two-point functions}
Let us denote a two-point function of the Carrollian primaries \eqref{primary} as 
\begin{equation}
    G^{(2)}(u_{1,2},\phi_{1,2})=\langle\mathcal O_{\Delta_1,\xi_1}(u_1,\phi_1)\mathcal O_{\Delta_2,\xi_2}(u_2,\phi_2)\rangle
\end{equation}
It is possible to completely determine the above expression by solving the Ward identities associated with the global BMS$_3$ generators, i.e.  $L_{0,\pm1}$ and $M_{0,\pm1}$. \\
The Ward identities associated with $M_0$ and $L_0$ allow $G^{(2)}$ only to be a function of the coordinate differences, i.e.
\begin{equation}
    G^{(2)}(u_{1,2},\phi_{1,2})\equiv  G^{(2)}(u_{12},\phi_{12}) 
\end{equation} 
Here $u_{12}=(u_1-u_2)$, $\phi_{12}=(\phi_1-\phi_2)$. The constraint from $M_1$ implies 
\begin{align}
   \( e^{i\phi_1}
\partial_{u_1}
 +e^{i\phi_2}
\partial_{u_2}
+i(\xi_1+\xi_2)\) G^{(2)}(u_{12},\phi_{12})=0
\end{align}
As usual this equation admits two classes of solutions. The so called \textit{magnetic branch} was previously studied in the literature and is given by 
\begin{align} \label{magnetic}
    G^{(2)}_{m}(u_{12},\phi_{12})=\left(
2\sin\frac{\phi_{12}}{2}
\right)^{-2\Delta}
\exp\!\left[
-\xi\,u_{12}\cot\frac{\phi_{12}}{2}
\right]\delta_{\Delta_1,\Delta_2}\delta_{\xi_1,\xi_2}.
\end{align}
 However there also exists ultralocal or \textit{electric branch} type of solutions of the form
\begin{equation} \label{inter}
     G^{(2)}(u_{12},\phi_{12})=f(u_{12})\delta(\phi_{12})\delta_{\xi_1+\xi_2,0}
\end{equation}
Finally the Ward identity associated with $L_1$ is given by
\begin{align}
     \( e^{i\phi_1}
(\partial_{\phi_1}+iu_1\p_{u_1}+i\Delta_1-\xi_1u_1)
 +e^{i\phi_2}
(\partial_{\phi_2}+u_2\p_{u_2}+i\Delta_2-\xi_2u_2\) G^{(2)}(u_{12},\phi_{12})=0
\end{align}
After plugging in \eqref{inter} in the above equation we obtain
\begin{align}
    iu_{12}\p_{u_1}f(u_{12})-\xi_1u_{12}f(u_{12})+i(\Delta_1+\Delta_2-1)f(u_{12})=0
\end{align}
is solved by
\begin{equation}
f(u_{12})=e^{i\xi_1u_{12}}u_{12}^{-(\Delta_1+\Delta_2-1)}
\end{equation}
This equation complete fixes the two point function as 
\begin{align} \label{two}
    G^{(2)}(u_{12},\phi_{12})=e^{i\xi_1u_{12}}u_{12}^{-(\Delta_1+\Delta_2-1)}\delta(\phi_{12})\delta_{\xi_1+\xi_2,0}
\end{align}
It is straightforward to check that this two point function naturally satisfies the constraints from $L_{-1}$ and $M_{-1}$. This \textit{electric branch} primary correlators  in presence of non-zero eigenvalue of $M_0$ has not been addressed in the literature before. 
\subsection{Three-point function}

Let´s denote the three point function as 
\begin{equation}
      G^{(3)}(u_{1,2,3},\phi_{1,2,3})=\langle\mathcal O_{\Delta_1,\xi_1}(u_1,\phi_1)\mathcal O_{\Delta_2,\xi_2}(u_2,\phi_2)\mathcal O_{\Delta_3,\xi_3}(u_3,\phi_3)\rangle
\end{equation}
In case of three-point functions, the electric branch itself admits a broad class of solutions. Here, however, we shall not be exhaustive and consider a special class of solutions with $\xi_1=\xi_2=0$ and $\xi_3=\xi \neq 0$. Let us start with the ansatz
\begin{align}
     G^{(3)}_{\Delta_i,\xi}(u_{1,2,3},\phi_{1,2,3})=F(\alpha_i(\phi_i),\phi_i)e^{i(\sum_{i=1}^3\alpha_i(\phi_i)u_i)}
\end{align}
$\alpha$ and $F$ are arbitrary functions, to be determined in the following. The invariance under $L_0$ imply
\begin{align}
\alpha_i(\phi_i+\theta)=\alpha_i(\phi_i), \quad F(\phi_i+\theta,\alpha_i(\phi_i+\theta))=F(\alpha_i(\phi_i),\phi_i)
\end{align}
Here $\theta$ denotes a constant angle. Furthermore, constraint arising from $M_0$ and $M_1$ are respectively given by
\begin{equation}
     \sum_{i}\alpha_i+i\xi=0,  \quad
     \sum_{i=1}^3 e^{i\phi_i}\alpha_i(\phi_i)=\xi e^{i\phi_3}
\end{equation}
The above equations are sufficient to completely determine the $\alpha_i(\phi_i)$s. They are given by
\begin{align}
\alpha_1=-\xi\frac{z_1z_{23}}{z_{12}.z_{13}},\quad \alpha_2=-\xi\frac{z_2z_{31}}{z_{12}.z_{23}}, \quad \alpha_3=\xi\frac{z_3^2-z_1 z_2}{z_{31}.z_{32}}
\end{align}
Here $z_{ij}=(z_i-z_j)$ and $z_i=e^{i\phi_i}$.
Furthermore the Ward identities from $L_{1}$ fixes $F(\alpha(z),z)$ as
\begin{equation} \label{3p}
    F(\alpha(z_i),z_i)=(z_1.z_2.z_3)\frac{\alpha_1^{\Delta_1-1}\alpha_2^{\Delta_2-1}\alpha_3^{\Delta_3-1}}{z_{12}.z_{23}.z_{31}}
\end{equation}
The generators $M_{-1}$ and $M_{-1}$ don't impose any further conditions.
Thus finally we have,
\begin{align}
     G^{(3)}_{\Delta_i,\xi}(u_{1,2,3},\phi_{1,2,3})=(z_1.z_2.z_3)\frac{\alpha_1^{\Delta_1-1}\alpha_2^{\Delta_2-1}\alpha_3^{\Delta_3-1}}{z_{12}.z_{23}.z_{31}}e^{i(\sum_{i=1}^3\alpha_i(\phi_i)u_i)}
\end{align}

\section{Massive scalar fields on 3d Minkowski spacetime} \label{3}

As advertised earlier, the correlation functions in previous section are expected to play a role in describing the massive scattering amplitudes in the bulk. In the following, we make these claims precise in a simple set up massive scalar fields in 3d Minkowski spacetime. We collect the relevant formulas before discussing a suitable extrapolation to $\mathcal{I}$. \\
A free massive scalar action is given by
\begin{align}
    \mathcal{S}=\int d^3x (\p_{\mu}\Phi\p^{\mu}\Phi+m^2\Phi^2)\
\end{align}
The Feynman propagator solves the equation of motion with a delta function source, i.e.
\begin{align} \label{eom}
    (\Box-m^2)G_F(X_1-X_2)=\delta^{(3)}(X_1-X_2)
\end{align}
 Here $X_i$ collectively denotes all the coordinates. Particularly for our purpose we would be using Bondi coordinates. The Minkowski metric in this coordinate is given by
 \begin{equation}
     ds^2=-du^2-2dudr+r^2d\phi^2
 \end{equation}
 In this coordinate, the Laplacian takes the following form 
 \begin{align} \label{laplacian}
    \Box=\left(
-2\,\partial_u\partial_r
+\partial_r^2
-\frac{2}{r}\partial_u
+\frac{1}{r}\partial_r
+\frac{1}{r^2}\partial_\phi^2
\right).
 \end{align}
 The solution to \eqref{eom} could be represented as 
 \begin{align} \label{Feynmann}
     G_F(X_1-X_2)=\int \frac{d^3p}{(2\pi)^3} \frac{1}{(p^2+m^2+i\epsilon)}e^{-ip.(X_1-X_2)}
,\end{align}
Here $p^{\mu}$ denotes the three momentum and on-shell satisfies the relation $p^2=-m^2$. It would be useful to consider the following parametrisation 
\begin{align} \label{massive par}
    p^\mu=(\omega+m)q^\mu(\theta)+im\p_{\theta}q^\mu(\theta), \quad q^{\mu}=(1,\cos \theta,\sin\theta)
\end{align}
This parametrisation generelises the standard parametrisation of massless momenta in celestial holography to satisfy the mass-shell condition $p^2=-m^2$.

\subsection{Bulk to boundary propagator}

Let us now discuss the extrapolation of this massive Feynmann propagator given in \eqref{Feynmann} to null infinity. The future null infinity ($\mathcal{I^+}$) is reached by taking $r \to \infty$, keeping $u,\phi$ fixed. Using the parametrisation in \eqref{massive par} and the Bondi coordinate, we can express the onshell propagator as 
\begin{align}
G_F(X_1-X_2)
&=
\frac{1}{2}
\int_{0}^{\infty} d\omega
\int_{0}^{2\pi} d\theta\,
\exp\Big[
i(m-\omega)
\Big\{
-u_{12}
-r_1\big(1-\cos\theta_1\big)
+r_2\big(1-\cos\theta_2\big)
\\ \nonumber
&\hspace{5.2cm}
+ m\big(
r_1\sin\theta_1
-r_2\sin\theta_2
\big)
\Big\}
\Big] .
\end{align}
Here we have used $u_{12}=u_1-u_2$ and $\theta_i=\theta-\phi_i$. We construct the bulk-to-boundary propagator by taking one of the bulk points, $X_1$, to null infinity. This is achieved by taking $r_1 \to \infty$, keeping the rest of the coordinates fixed. Notice that, in this limit, the above expression becomes rapidly oscillating for real values of the mass parameter $m$, and hence localises the $\theta$ integral via the condition $\theta_1=0$. In this limit the leading order term behaves as $r_1^{-\frac{1}{2}}$, after rescaling we obtain,
\begin{align} \label{bulk to bdy}
\psi(X_2;u_1,\phi_1)&=\int_{0}^{\infty} d\omega \exp\Big[
i(m-\omega)
\Big\{
-u_{12} +r_2\big(1-\cos{(\phi_1-\phi_2)}\big)+imr_2\sin{(\phi_1-\phi_2)}\Big] \\ \nonumber
&=\exp\Big[im\Big(-u_{12}+r_2(1-\cos{(\phi_1-\phi_2)})-ir_2\sin{(\phi_1-\phi_2)}\Big)\Big] \\ \nonumber
& \hspace{3.2cm} \times \int_{0}^{\infty}d\omega \omega^{\frac{1}{2}} \exp{\Big[-i\omega\Big(-u_{12}+r_2(1-\cos{(\phi_1-\phi_2)}\Big)\Big]} \\ \nonumber
&= \frac{\exp\Big[im\Big(-u_{12}+r_2(1-\cos{(\phi_1-\phi_2)})-ir_2\sin{(\phi_1-\phi_2)}\Big)\Big]}{\Big(-u_{12}+r_2(1-\cos{(\phi_1-\phi_2)}\Big)^{\frac{1}{2}}}
\end{align}
The above expression defines the bulk-to-bulk to boundary propagator for a massive scalar field $\Phi(u_2,r_2,\phi_2)$ and the corresponding boundary operator $\mathcal{O}_{\Delta,\xi}(u_1,\phi_1)$ at null infinity of operator dimension $\Delta=\frac{1}{2}$ and $\xi=m$. It is straightforward to check that the above expression indeed reduces to the bulk to the boundary propagator of massless fields if we set $m=0$, previously derived in \cite{Bagchi:2023fbj,Alday:2024yyj}. \\

\subsection{Carrollian primary wavefunction} \label{priwave}

 The bulk-to-boundary propagator serves the purpose of an intertwiner between the representations of the bulk isometry and the boundary conformal isometry group. A bulk field $\Phi(X)$ of mass $m$ and spin $j$ could be expressed in terms of the corresponding boundary operator $\mathcal{O}_{\Delta,\xi}(u,\phi)$ as 
\begin{align}
    \Phi(X)=\int dud\phi\psi_{\Delta,\xi}(X;u,\phi)\mathcal{O}_{\Delta,\xi}(u,\phi)
\end{align}
The above equation requires the integral kernel $\psi_{\Delta,\xi}$ to transform in the following way by the simultaneous action of the bulk Poincaré $(\Lambda, a)$ and the associated boundary conformal Carroll transformations $f(\phi),\alpha(\phi)$.
\begin{align}
    \psi_{\Delta,\xi}(\Lambda X+a;u',\phi')
=
\mathcal{J}_{\Delta,\xi}(\Lambda,a;u,\phi)
\psi_{\Delta,\xi}(X;u,\phi)
\end{align}
Where 
\begin{equation*}
    \mathcal{J}_{\Delta,\xi}(\Lambda,a;u,\phi)=\big(f'(\phi)\big)^{\Delta}
\exp\!\left(
\xi\,\frac{\alpha'(\phi)+\dfrac{uf''(\phi)}{f'(\phi)}}
{f'(\phi)}
\right), \quad \phi'=f(\phi), \quad
u'=f'(\phi)(u+\alpha(\phi))
\end{equation*}
The above equation is the Carrollian analog of  \textit{conformal primary wavefunction} introduced in \cite{Pasterski:2016qvg}, which substitutes the Lorentz group and a point on Celestial sphere  respectively by the Poincare group and a point on null infinity.\\

In 3D, these transformation rules
 can be expressed as six differential equations associated with each global generator. These equations are sufficient to entirely determine the functional form of $\psi_{\Delta,\xi}$. The equations are given by
 \begin{align}
\left(P_n^{\rm bulk}+M_n^{\rm bdry}\right)\psi_{\Delta,\xi}=0,
\qquad
\left(J_n^{\rm bulk}+L_n^{\rm bdry}\right)\psi_{\Delta,\xi}=0,
\qquad n=0,\pm1.
 \end{align}
Let us introduce
\begin{align}
    S=u-t+x\cos\phi+y\sin\phi, \quad
R=-x\sin\phi+y\cos\phi
\end{align}

\begin{itemize}

\item \textbf{The equation from $M_0-P_0$}
\begin{equation}
    (P_t+M_0)\psi=0 \implies (\partial_t+\partial_u)\psi=0.
\end{equation}
Therefore,
\begin{equation}
    \psi_{\Delta,\xi}=\psi_{\Delta,\xi}(u-t,x,y,\phi)
\end{equation}
\item \textbf{The equation from $L_0-J_0$ equation}
\[
(x\p_y-y\p_x+\p_{\phi})\psi=0.
\]
Since
\begin{equation}
    (x\p_y-y\p_x+\partial_\phi)S=0,
\qquad
(x\p_y-y\p_x+\partial_\phi)R=0
\end{equation}

the kernel can only depend on $X$ through $S$ and $R$, i.e. $\psi=\psi(S,R)$.

\item \textbf{The equation from $M_1-P_{+}$ equation}
\[
(\partial_x+i\partial_y-e^{i\phi}(\partial_u+i\xi))\psi=0
\]
Now
\[
(\partial_x+i\partial_y)S=e^{i\phi},
\qquad
(\partial_x+i\partial_y)R=i\,e^{i\phi}.
\]

Using the above expressions, we can write,
\[
(\partial_x+i\partial_y)\psi
=
e^{i\phi}
\left(
\partial_S+i\partial_R
\right)\psi,
\]
while
\[
e^{i\phi}(\partial_u+i\xi))\psi 
=
e^{i\phi}
\left(
\partial_S+i\xi
\right)\psi.
\]

Hence
\[
((\partial_x+i\partial_y)-e^{i\phi}(\partial_u+i\xi))\psi
=
i\,e^{i\phi}
(\partial_R-\xi)\psi=0,
\]
which implies
\[
\psi(S,R)=e^{\xi R}F(S).
\]
\item \textbf{The equation from $L_1-J_{+}$ equation}
\[
((y-ix)\partial_t
+
t(\partial_y-i\partial_x)+e^{i\phi}
\left(
\partial_\phi
+i u\partial_u
+i\Delta
-\xi u
\right))\psi=0
\]
After plugging in the expression we obtained from $M_1$, we obtain
\[
S F'(S)+\Delta F(S)+i\xi S F(S)=0.
\]
Integrating this equation gives
\[
F(S)
=
C_{\Delta,\xi}
S^{-\Delta}
e^{-i\xi S}.
\]
\end{itemize}
The equations from $L_1$ and $M_1$ are satisfied by the above expression, but do not constrain it any further. Finally, we have \footnote{Following this line of arguments, it is analogously possible to determine the bulk-to-boundary propagator in AdS/CFT. See Appendix \ref{A} for a detailed derivation. }
\begin{align}
    \psi_{\Delta,\xi}(X;u,\phi)=C_{\Delta,\xi}
S^{-\Delta}
e^{\xi(R-iS)}
\end{align}
This expression of the \textit{Carrollian primary wavefunction} indeed agrees with the expression of the scalar bulk-to-boundary propagator in \eqref{bulk to bdy} if we set $\Delta=\frac{1}{2}$. \\

The \textit{Carrollian primary wavefunction} admits the following integral representation 
\begin{align} \label{basisc}
    \psi_{\Delta,\xi}(X;u,\phi)=\int _{0}^{\infty}d\omega \omega^{\Delta-1}e^{i(\omega+\xi)u}e^{ip(\omega,\phi).X}
\end{align}
Here, $e^{ipx}$ represents the standard momentum eigenbasis for massive particles in Minkowski spacetime. Using this expansion in plane wave basis and the mass-shell condition,  it is straightforward to verify that $\psi_{\Delta,\xi}$ satisfies the massive Klien-Gordon equation without any source, i.e.
\begin{equation}
    (\Box-m^2)\psi_{\Delta,\xi}(X;u,\phi)=0
\end{equation}

\section{ Boundary correlators and scattering amplitudes} \label{4}

In this section, we explicitly compute flat space analog of elementary Witten diagrams using the propagators evaluated in the previous section.
These Witten diagrams would result to the correlation functions fixed using symmetry arguments. 
\subsection{Two-point diagram}

Let us consider the simplest case of a two-point contact diagram. This is given by
\begin{align}
   \mathcal{A}_2(u_1,\phi_1;u_2,\phi_2)=\int d^3x \psi_{\xi_1} (x;u_1,\phi_1)\psi_{\xi_2} (x;u_2,\phi_2)
\end{align}

Plugging in the integral representation of $\psi_{\xi} (x;u,\phi)$ in \eqref{basisc}, we end up with
\begin{align}
    \mathcal{A}_2(u_1,\phi_1;u_2,\phi_2)=&\int d^3x \int_0^{\infty} d\omega_1 \omega_1^{\Delta_1-1}\int_{0}^{\infty} d\omega_2 \omega_2^{\Delta_2-1} \\ \nonumber
    & \times e^{i((\omega_1+\xi_1)u_1+(\omega_2+\xi_2)u_2)}e^{i(p_1(\omega_1,\phi_1)+(p_2(\omega_2,\phi_2)).X}
\end{align}
We perform the x-integral first, which yields momentum conservation. 
\begin{align} \mathcal{A}_2(u_1,\phi_1;u_2,\phi_2)=\int_0^{\infty} d\omega_1 \omega_1^{\Delta_1-1}\int_{0}^{\infty} d\omega_2 \omega_2^{\Delta_2-1} e^{i((\omega_1+\xi_1)u_1+(\omega_2+\xi_2)u_2)}\delta^{(3)}(p_1+p_2)
\end{align}
For a fixed value of $\xi$, the momentum conservation implies
\begin{equation}
p_1+p_2=0
\quad\Longrightarrow\quad
\omega_1=-\omega_2,
\quad
\phi_1=-\phi_2.
\end{equation}
After performing the $\omega$ integrals, taking into account the above-mentioned input, we finally arrive at
\begin{align}
\mathcal{A}_2(u_1,\phi_1;u_2,\phi_2)
=u_{12}^{-(\Delta_1+\Delta_2-1)}e^{-i\xi u_{12}}\delta(\phi_{12})
\end{align}
The above expression indeed agrees with the two-point function of Carrollian primaries \eqref{two}.

\subsection {Three-point contact diagram}

Here, we introduce a cubic vertex among the massive scalar with an additional massless scalar field and evaluate a contact three point Witten diagram. The interaction term is given by
\begin{align}
    \lambda\int d^3x \Phi^2\Phi_m
\end{align}
Where, $\lambda$ is the coupling constant and $\Phi_m$ and $\Phi$ respectively denote the massive and the massless fields.A three point contact diagram is given by
\begin{equation}
{\cal A}_3 (u_i,\phi_i)
=
\lambda
\int d^3x\,
\psi_{\Delta_1,0}(x;u_1,\phi_1)
\psi_{\Delta_2,0}(x;u_2,\phi_2)
\psi_{\Delta_3,\xi}(x;u_2,\phi_3).
\end{equation}
Substituting the Schwinger representation gives
\begin{align}
\mathcal{A}_3{(u_i,\phi_i)}
=
\lambda
\int_0^\infty
&\prod_{i=1}^3 d\omega_i \omega_1^{\Delta_1-1}\omega_2^{\Delta_2-1}\omega_3^{\Delta_3-1}\,
\nonumber\\
&\times
e^{i\omega_1u_1}
e^{i\omega_2u_2}
e^{+i(\omega_3+\xi)u_3}
\int d^3x\,
e^{-i(p_1+p_2-p_3)\cdot x}.
\end{align}

The bulk integral gives

\begin{equation}
\int d^3x\,
e^{i(p_1+p_2+p_3)\cdot x}
=
(2\pi)^3
\delta^{(3)}(p_1+p_2+p_3).
\end{equation}

In components, the momentum conservation constraints become
\begin{align}
C_0
&=
\omega_1+\omega_2+\omega_3+\xi=0,
\\ \nonumber
C_+
&=
\omega_1 z_1+\omega_2 z_2+\omega_3 z_3=0,
\\ \nonumber
C_-
&=
\omega_1 z_1^{-1}
+\omega_2 z_2^{-1}
+(\omega_3+2\xi)z_3^{-1}=0.
\end{align}
Here, we have used $z_i=e^ {i\phi_i}$ as a shorthand. $C_{\pm}$ are the conservations of $p_{\pm}=p_x \pm ip_y$ components.
Solving the constraints yields
\begin{align}
\omega_1^\star=\xi
\frac{z_1(z_2-z_3)}
{(z_1-z_2)(z_1-z_3)}
, \quad
\omega_2^\star=\xi
\frac{z_2(z_3-z_1)}
{(z_1-z_2)(z_2-z_3)}
, \quad
\omega_3^\star=-\xi
\frac{
z_3(z_1+z_2)-2z_1z_2
}
{(z_3-z_1)(z_3-z_2)}
\end{align}

The Jacobian is given by

\begin{equation}
J=
\det
\frac{\partial(C_0,C_+,C_-)}
{\partial(\omega_1,\omega_2,\omega_3)}
=
\frac{
(z_1-z_2)(z_1-z_3)(z_2-z_3)
}
{z_1z_2z_3}.
\end{equation}

Performing the $\omega_i$ integrals gives

\begin{equation}
{\cal A}_3(u_i,\phi_i)
=
\frac{{\cal N}_3}{J}{\omega_1^\star}^{\Delta_1-1}{\omega_2^\star}^{\Delta_2-1}{\omega_3^\star}^{\Delta_3-1}
e^{i\omega_1^\star u_1}
e^{i\omega_2^\star u_2}
e^{i(\omega_3^\star+\xi)u_3}.
\end{equation}

This contact diagram indeed agrees with the three-point correlation function derived previously in \eqref{3p}. \\

We conclude this section by establishing a formula that relates the 2d conformal Carroll correlator to 3d scattering amplitudes. This was achieved previously for the massless case by going into the so-called \textit{Carrollian primary} basis. In the previous sections, we have extended the notion of Carroll primary wavefunctions with an additional quantum number that captures the mass in the bulk. 
Notice that the resulting integral transform in \eqref{basisc} that relates the momentum eigenbasis and the Carrollian primary wavefunction in this case is a generalisation of the so-called \textit{Modified Mellin transformation} \cite{Banerjee:2018gce, Banerjee:2019prz}. This generalised integral transformation maps the momentum space scattering amplitudes to the standard form of boundary Carroll correlators. To be precise, an n-point massive amplitude $\mathcal M_n(n_{in}\to n_{out})$, is related to an n-point conformal Carroll correlation function by
\begin{equation}
\boxed{
\begin{aligned}
\left\langle \prod_{i=1}^{n}
\mathcal O_{\Delta_i,\xi_i}(u_i,\phi_i)
\right\rangle
&=
\int_0^\infty
\prod_{a=1}^{n_{\rm in}}
\left[
d\omega_a\,
\omega_a^{\Delta_a-1}
e^{+i(\omega_a+\xi_a)u_a}
\right]
\\
&\quad\times
\prod_{b=1}^{n_{\rm out}}
\left[
d\omega_b\,
\omega_b^{\Delta_b-1}
e^{-i(\omega_b+\xi_b)u_b}
\right]
\\
&\quad\times
\mathcal M_n
\!\left(
p^{\rm in}_1,\ldots,p^{\rm in}_{n_{\rm in}};
p^{\rm out}_1,\ldots,p^{\rm out}_{n_{\rm out}}
\right).
\end{aligned}
}
\end{equation}

\medskip

\section{Discussions}

In this paper, we generalized the framework of Carrollian amplitudes to incorporate massive fields in the bulk. We considered a setup of scalar fields in three-dimensional Minkowski space. We revisited aspects of representations of 2d Carrollian CFTs and verified the existence of the electric branch of correlators in a generalised case. We constructed a bulk-to-boundary propagator by suitably extrapolating the Feynman propagator of a massive scalar field in three dimensions to null infinity. This bulk-to-boundary propagator generalises the Carrollian primary wavefunctions to incorporate a new quantum number related to the mass in the bulk. Finally, we showed that the scattering amplitudes involving massive particles take the form of Carrollian correlation functions derived before. \\

One immediate direction for the future is to understand this construction in higher dimensions. The primary fields defined in higher dimensional Carrollian CFTs are usually labelled by the scaling dimensions and the spin \cite{Bagchi:2016bcd}.
\begin{equation*}
    [D,\mathcal{O}(0,0)]=\Delta\mathcal{O}(0,0), \quad [J,\mathcal{O}(0,0)]=\sigma\mathcal{O}(0,0)
\end{equation*}
However, the quadratic Casimir of Poincare group is related to conformal Carroll generators via
 \begin{equation*}
   P^2=B_iB^i-HK
\end{equation*}
Where, $B_i,H$ and $K$ are respectively Carroll boost, Hamiltonian and the special conformal transformation. One can check that the above representation of Carrollian primaries would correspond to massless representation of Poincare in the bulk. In order to construct the massive representations it is essential to label these primary operators with one of the generators that contribute to the Casimir.\\

The incorporation of mass also gives a window to explore other asymptotically flat spacetimes. The asymptotically flat solutions of 3d Einstein gravity contains massive solutions \cite{Barnich:2006av}. These spacetimes can be generated by taking backreaction of the massive particles into account. The boundary correlation function of probe fields in these background would correspond to higher point functions because of the insertions of heavy operators accounting for the heavy background \cite{Hijano:2015qja,Fitzpatrick:2015zha}. The notion of BMS$_3$ heavy-light higher point functions would perhaps be useful in addressing scattering amplitudes in non-trivial asymptotically flat backgrounds.

\medskip

\subsection*{Acknowledgement}

It is a pleasure to thank Arjun Bagchi, Shamik Banerjee, Robinson Mancilla and Jakob Salzer for helpful discussions. We also thank the organisers and the participants of the Solvay workshop on \textit{Carrollian physics and geometry}, where this article was presented.
This work is supported by Fonds de la Recherche Scientifique F.R.S.-
FNRS (Belgium) through the project PDR/OL C62/5 “Black hole horizons: away from
conformality” (2022-2025) and a grant from the International Solvay Institute.

\bigskip

\newpage

\begin{appendices}

\section{The bulk-to-boundary propagator in AdS$_3$ using symmetry arguments} \label{A}

Consider AdS$_3$ in Poincar\'e coordinates,
\begin{equation}
ds^2=\frac{dz^2+dx^+dx^-}{z^2}.
\end{equation}

Let $K(z,x^\pm;y^\pm)$ denote the bulk-to-boundary propagator between a bulk point
$(z,x^+,x^-)$ and a boundary point $(y^+,y^-)$.

The isometry group of AdS$_3$ is
\[
SO(2,2)\simeq SL(2,\mathbb R)_L\times SL(2,\mathbb R)_R.
\]
The kernel transforms as a bulk scalar and as a boundary primary of weights
$(h,\bar h)$. Therefore
\[
(L_n^{\rm bulk}+L_n^{\rm bdry})K=0,
\qquad
(\bar L_n^{\rm bulk}+\bar L_n^{\rm bdry})K=0,
\]
for $n=0,\pm1$.

\subsection*{Differential equations :}

The left-moving generators acting on the bulk are
\begin{align}
L_{-1}^{\rm bulk}&=\partial_{x^+},\\
L_{0}^{\rm bulk}&=x^+\partial_{x^+}+\frac{z}{2}\partial_z,\\
L_{1}^{\rm bulk}&=(x^+)^2\partial_{x^+}+x^+z\partial_z-z^2\partial_{x^-},
\end{align}
while on a boundary primary of weight $h$,
\begin{align}
L_{-1}^{\rm bdry}&=\partial_{y^+},\\
L_{0}^{\rm bdry}&=y^+\partial_{y^+}+h,\\
L_{1}^{\rm bdry}&=(y^+)^2\partial_{y^+}+2hy^+.
\end{align}
Analogously there are three more right-moving generators. The differential equations imposed by these generators are given by the following equations
\begin{align}
&(\partial_{x^+}+\partial_{y^+})K=0, \label{eq1}\\
&(\partial_{x^-}+\partial_{y^-})K=0,\label{eq2}\\
&\left(x^+\partial_{x^+}+\frac z2\partial_z
+y^+\partial_{y^+}+h\right)K=0,\label{eq3}\\
&\left(x^-\partial_{x^-}+\frac z2\partial_z
+y^-\partial_{y^-}+\bar h\right)K=0,\label{eq4}\\
&\left((x^+)^2\partial_{x^+}+x^+z\partial_z-z^2\partial_{x^-}
+(y^+)^2\partial_{y^+}+2hy^+\right)K=0,\label{eq5}\\
&\left((x^-)^2\partial_{x^-}+x^-z\partial_z-z^2\partial_{x^+}
+(y^-)^2\partial_{y^-}+2\bar h\,y^-\right)K=0.\label{eq6}
\end{align}
Equations (\ref{eq1}) and (\ref{eq2}) imply that the kernel depends only on coordinate differences,
\[
u=x^+-y^+,
\qquad
v=x^--y^-,
\]
so that
\[
K(z,x^\pm;y^\pm)=F(z,u,v).
\]

Since
\[
\partial_{x^+}=\partial_u,
\qquad
\partial_{y^+}=-\partial_u,
\]
and similarly for $v$, equations (\ref{eq3}) and (\ref{eq4}) become
\begin{align}
\left(
u\partial_u+\frac z2\partial_z+h
\right)F&=0,\label{dil1}\\
\left(
v\partial_v+\frac z2\partial_z+\bar h
\right)F&=0.\label{dil2}
\end{align}

For a scalar field,
\[
h=\bar h=\frac{\Delta}{2},
\]
and one finds that
\begin{equation}
F(z,u,v)=z^{-\Delta}f(\eta),
\qquad
\eta=\frac{uv}{z^2}.
\end{equation}
Equation (\ref{eq5}) becomes
\begin{equation}
\left(
u^2\partial_u
+uz\partial_z
-z^2\partial_v
+2hu
\right)F=0,
\end{equation}
while (\ref{eq6}) gives
\begin{equation}
\left(
v^2\partial_v
+vz\partial_z
-z^2\partial_u
+2\bar h\,v
\right)F=0.
\end{equation}

Substituting
\[
F=z^{-\Delta}f(\eta),
\qquad
\eta=\frac{uv}{z^2},
\]
one computes
\begin{align}
\partial_uF&=
z^{-\Delta}\frac{v}{z^2}f'(\eta),\\
\partial_vF&=
z^{-\Delta}\frac{u}{z^2}f'(\eta),\\
\partial_zF&=
-\Delta z^{-\Delta-1}f(\eta)
-2z^{-\Delta-1}\eta f'(\eta).
\end{align}

Substituting into the first special conformal equation yields
\[
(1+\eta)f'(\eta)+\Delta f(\eta)=0.
\]

The second equation gives exactly the same condition.

Integrating,
\[
\frac{f'(\eta)}{f(\eta)}
=
-\frac{\Delta}{1+\eta},
\]
so that
\[
f(\eta)=C(1+\eta)^{-\Delta}.
\]

Therefore
\begin{align}
K(z,x^\pm;y^\pm)
&=
C\,z^{-\Delta}
\left(
1+\frac{uv}{z^2}
\right)^{-\Delta}
\\
&=
C
\left(
\frac{z}{z^2+uv}
\right)^\Delta.
\end{align}

Restoring $u=x^+-y^+$ and $v=x^--y^-$, one finally obtains

\begin{equation}
\boxed{
K_\Delta(z,x^\pm;y^\pm)
=
C_\Delta
\left[
\frac{z}
{z^2+(x^+-y^+)(x^--y^-)}
\right]^\Delta
}
\end{equation}

which is the scalar bulk-to-boundary propagator in AdS$_3$.
\end{appendices}

\newpage


\newpage

\bibliographystyle{JHEP}
\bibliography{flat}

\end{document}